\documentstyle[12pt,aasms]{article}
\begin{document}

\title{NGC 4314. III.\ Inflowing Molecular Gas Feeding \\ a
Nuclear  Ring of Star Formation\footnote{Based on observations
collected at the Owens Valley  Radio Observatory}}

\author{G.\ Fritz Benedict}
\affil{McDonald Observatory, University of Texas, Austin, TX
78712}

\author{Beverly J.\ Smith}
\affil{IPAC, California Institute of Technology, Pasadena, CA
91125}
\and 
\author{Jeffrey D.\ P.\ Kenney}
\affil{Astronomy Dept., P.O.\ Box 208101, Yale University, New
Haven, CT 06520}

\begin{abstract}

NGC 4314 is an early-type barred galaxy containing a nuclear ring
of  recent star formation. We present CO(1-0) interferometer data
of the bar and  circumnuclear region  with $2.3^{\prime \prime}
\times 2.2^{\prime \prime}$  spatial resolution and 13  km
s$^{-1}$ velocity resolution acquired at the Owens Valley  Radio
Observatory . These data reveal a  clumpy circumnuclear ring of
molecular gas. We also find a peak of CO inside  the ring within
2$^{\prime \prime}$ of the optical center that is  not associated
with massive star  formation. We construct a rotation curve from
these CO kinematic data and the  mass model of Combes et al.\
(1992). Using this rotation curve, we have identified  the
location of orbital resonances in the galaxy. Assuming that the
bar ends at  corotation, the circumnuclear ring of star formation
lies between two Inner  Lindblad Resonances, while the nuclear
stellar bar ends near the IILR. Deviations  from circular motion
are detected just beyond the CO and H$\alpha$ ring, where the 
dust lanes along the leading  edge of the bar intersect the
nuclear ring. These non-circular motions along the minor axis
correspond to radially inward streaming  motions at speeds of $20
- 90$ km s$^{-1}$ and clearly show inflowing gas feeding an  ILR
ring. There are bright H\,{\sc ii} regions near the ends of this
inflow region,  perhaps indicating triggering of star formation
by the inflow.

\end{abstract}

\section{Introduction}

Inner Lindblad resonances (ILRs) are thought to strongly
influence  gas motions and star formation in the central regions
of galaxies.  Gas  surface densities often peak near the ILR
(Combes et al.\ 1992; Kenney et  al.\ 1992), lending support to
the  idea that the inward flow of gas along  galaxian bars slows
down and piles up between the OILR and the IILR  (Combes 1988;
Shlosman et al.\ 1989).  Studies of the CO (1-0) distribution
within the central regions of barred spirals (Kenney et al.\ 
1992; Kenney 1995; Regan, Vogel, \& Teuben 1995; Rand 1995; 
Sakamoto et al.\ 1995) show a variety of molecular gas
morphologies near  the ILR, including rings, spiral arms, and
``twin peaks'' with two maxima  where the dust lanes along the
primary bar cross the ILR.  In  the case of  the starburst galaxy
NGC 3504, the CO peaks near the nucleus, probably  inside the ILR
(Kenney et al.\ 1993).  The dynamical and evolutionary 
differences responsible for this variety are not yet understood,
but are  critical for understanding starbursts and the formation
and evolution of  compact circumnuclear stellar disks (Kormendy
1993) and stellar bulges  (Pfenniger \& Norman 1990).
	
At present, only a handful of barred spirals have been mapped in
CO  at sufficient resolution to study the effects of ILRs on the
gas distribution and kinematics. The barred spiral galaxy NGC
4314 is a particularly good galaxy to study because it exhibits
evidence for many features  associated with resonances and it is
nearly face-on ($i \sim 23\deg$). It has a large-scale stellar
bar of diameter 130$^{\prime \prime}$ (6.3 kpc) and a prominent
circumnuclear  ring of  star formation of diameter 10$^{\prime
\prime}$  (500 pc) that is visible in H$\alpha$  (Pogge  1989),
radio continuum (Garcia-Barreto et al.\ 1991; hereafter G-B),
and  optical color maps (Benedict et al.\ 1992; hereafter P1). CO
(1-0) mapping  of NGC 4314 (Combes et al.\ 1992) at 5$^{\prime
\prime}$  resolution revealed the presence   of a molecular ring
slightly smaller than that of the ring of star formation. 
Outside of these rings, a blue elliptical feature of diameter
20-25$^{\prime 
\prime}$  (1 kpc)  is seen in optical color maps, which may
correspond to a ring of relatively  young but non-ionizing stars
(P1).  This feature, which can be seen on the  unsharp masked
optical image shown in Fig.\ 1 (from P1), is elongated 
perpendicular to the primary bar, suggesting that it extends
outward to the  IILR. Combes et al.\ (1992) suggest that in this
galaxy star formation is  propagating inwards. This is in
contrast with the evolutionary scenario  proposed by Kenney et
al.\ (1993) for strong starbursts, in which star  formation
devours gas most rapidly in the center, ultimately forming a 
ring of gas near the ILR.
	
Inside the H\,{\sc ii} region ring, a nuclear stellar bar of
diameter 8$^{\prime 
\prime}$  (400  pc) is seen in {\it Hubble Space Telescope} (HST)
I band data (Benedict et  al.\  1993; hereafter P2). This nuclear
bar is aligned parallel with the large  scale bar, which suggests
that the nuclear bar has the same pattern speed as  the primary
bar, and lies within the IILR (Binney \& Tremaine 1987).

In this paper, we present new CO (1-0) maps of NGC 4314,
obtained  at twice the spatial resolution as the Combes et al.\
(1992) data. The  observations are described in Section 2, while
the CO distribution and  kinematics are discussed in Sections 3
and 4, respectively.  In Section 5 we  compare NGC 4314 with
other galaxies and discuss star  formation in this galaxy. Some
general properties of NGC 4314 are  tabulated in Table 1.

\section{Observations}

We observed NGC 4314 in the CO (1-0) line using the Owens Valley
Radio  Observatory (OVRO) millimeter-wave interferometer (Padin
et al.\ 1991) in five  configurations between April and June
1991. Projected baselines ranged in length  from 10 to 200
meters. At that time OVRO consisted of a three element array 
with 10.4m dishes (HPBW = 65$^{\prime \prime}$  and SIS
receivers. The 32 5  MHz receivers  give 13 km s$^{-1}$
resolution and an instantaneous bandpass of 416 km s$^{-1}$. The 
quasar $1219 + 285$ was used for phase calibration, Uranus and
Neptune for flux  calibration. 

After calibration, channel maps were made with both uniform and
natural  weightings using AIPS. Uniform weighting produces higher
resolution maps,  with less sensitivity, while natural weighting
provides greater sensitivity with less  resolution. The latter
reduction mode provides higher signal to noise information  on
the regions in the primary bar of NGC 4314 containing fainter CO
intensity. 

All maps were made as $256 \times 256$ arrays with 0\farcs5
pixel$^{-1}$. Natural  weighting produced a resolution of $3.1
\times 2\farcs7$ with channel noise 0.022 Jy  beam$^{-1}$. These
maps were CLEANed to a 0.045 Jy beam$^{-1}$ level. Using 
uniform  weighting, we obtained a $2.3 \times 2\farcs2$ beam with
channel noise 0.028 Jy  beam$^{-1}$,  CLEANed to a 0.07 Jy
beam$^{-1}$ level. For both weightings we have made primary  beam
corrections. Fig.\ 2 presents our uniformly weighted channel
maps. These  maps have absolute positional accuracy of about
0\farcs5. Table 2 summarizes the  interferometer-related
parameters.	

The total CO flux observed by OVRO is 245 Jy km s$^{-1}$ (Table
2), while the  CO luminosity is $4.7 \times 10^7$ K km s$^{-1}$
pc$^{2}$. This luminosity is consistent with  the  single dish
observations of G-B and with the Nobeyama Millimeter Array
results  of Combes et al.\ (1992) (see Table 1). Therefore we
have recovered essentially  all of the CO flux from this galaxy
with our OVRO observations.

\section{Morphological Results}

\subsection{CO Morphology}

Uniformly weighted and naturally weighted total intensity maps
were  constructed using the AIPS moment routines.  Fig.\ 3
presents the uniformly  weighted CO intensity map. The strongest
CO emission arises from an  incomplete, clumpy ring with an
off-center minimum. There are five clumps of  emission along the
ring, separated by gaps. The largest of these gaps is at p.a.\ = 
239\deg. A secondary gap occurs at p.a.\ = 330\deg. In
addition to the clumps in the  ring, we find strong CO emission
within 2$^{\prime \prime}$  of the optical center. 
	
Two symmetrically located features along p.a. $\sim$20\deg\
extend out from  the  ring. These are associated with the dust
lanes that lie along the leading edge of the  primary stellar
bar. These lanes are detected in CO as they curve around to 
intersect the nuclear ring. We refer to these features as inflow
spurs, since the  CO velocity field shows clear evidence for
inflow motions here (\S4.2).

Fig.\ 4 shows the naturally weighted CO intensity distribution,
covering  about 4 times the area on the sky as Fig.\ 3. Note the
CO clumps extending on  either side out from the ring along p.a.\
$\sim$148\deg.   These extensions lie along the  leading edge
of the primary stellar bar (P1). 

\subsection{Comparison of CO to Optical and Radio Continuum}

\subsubsection{Maps}

In Fig.\ 5a we overlay the uniform-weight CO map contours on the
optical  unsharp masked image shown in Fig.\ 1.  The dust and CO
are relatively  coincident for the CO arcs to the SE and NW,
while the dust lanes lie outside the  CO ring to the NE and SW.
Along the primary bar, a dusty area is seen to the SE  (see Fig.\
1), a region called the `dust bowl' in P1. A comparison of
optical and  CO intensity from the naturally-weighted data for a
larger area of the galaxy  (Fig.\ 5b) shows weak ($\sim 3\sigma$)
CO sources coincident with  the dust bowl. Note in  both Fig.\ 5a
and 5b the anticoincidence of CO with the outer ellipse of
relatively  young stars.

In Fig.\ 6 the Pogge (1989) H$\alpha$ + [N II] map is overlaid on
a uniform weight   CO intensity contour plot. The Wakamatsu \&
Nishida (1980) H$\alpha$-bright knots  A  (p.a.\ $\sim$
90\deg) and B (p.a.\ $\sim$ 178\deg) are the second and
third strongest H$\alpha$  sources in  the Pogge map. The
H$\alpha$ peaks tend to lie  outside of the CO ring (see also
the  profiles in Fig.\ 7 - 9). At the largest CO gap,  H$\alpha$
is also weak.

The H$\alpha$ + [N II] map resembles the 6 cm radio continuum map
shown in   G-B and our observations confirm the results of Combes
et al.\ (1992) that the ring  of star formation has a larger
radius than the molecular ring.

\subsubsection{Profiles}

The profiles in Fig.\ 7 - 10 offer a more detailed and
quantitative  comparison between the distributions of CO
intensity, blue light ($\mu_{\rm B}$, P1),  
$\mu_{({\rm H}\alpha + [{\rm N} II])}$ (uncalibrated H$\alpha$+[N
II] surface  intensity, Pogge 1989), $\mu_{\rm I-J}$ and 
$\mu_{\rm B-H}$ (respectively identified as the best tracer of
dust and of new stars in P1).  Fig. 7, 8, and 9 are primary bar
major axis, primary bar minor axis, and E-W  cuts across the
galaxy, respectively. Fig. 10 shows a cut through the dust bowl, 
perpendicular to the primary bar.  For these plots the
uniformly-weighted CO  intensity has been converted to
$\sigma_{\rm H_{2}}$(M$_{\sun}$ pc$^{-2}$)  assuming (Bloeman et
al.\ 1986,  Kenney et al.\ 1992)
\begin{displaymath} 
\sigma_{{\rm H}_2} ({\rm M}_{\sun}{\rm pc}^{-2}) = 470 \;{\rm
I}_{\rm CO}\;  ({\rm Jy\, km\, s}^{-1} {\rm arcsec}^{-2})
\end{displaymath}  H$\alpha$ is a tracer of the local star
formation rate (SFR). These plots also contain the  log of the
ratio of (H$\alpha$+[N II])/CO. In H\,{\sc ii}  regions, where
the observed H$\alpha$+[N II]  emission is dominated by H$\alpha$
and the  ionization is due to OB stars, this ratio is 
proportional to the gas depletion timescale, or Star Formation
Efficiency. 
	
In Fig.\ 7, the dust lanes at $r = \pm 5\farcs5$, seen as
reddening in the
$\mu_{\rm I-J}$ profile,  are very prominent in CO. There is also
considerable CO  emission near the  nucleus; Fig.\ 7, 8, and 9
show the CO surface density near the nucleus as high as  in the
ring of star formation. The $\mu_{({\rm  H}\alpha+[{\rm N} II])}$
profiles corroborate that the H\,{\sc ii}  regions in the ring
lie slightly outside the CO ring.  Dips in the $\mu_{\rm B-H}$
profile are  coincident with the new stars  at the  H$\alpha$
ring, $r = \pm 7^{\prime \prime}$ . The (H$\alpha$+[N II])/CO
ratio  shows a relatively  higher SFE near the nuclear ring,
especially to the NW. 
	
There is also a secondary source in  $\mu_{({\rm H}\alpha + [{\rm
N\, II}])}$,  peaking $\sim1^{\prime \prime}$  S of the  center.
This is likely due to strong [N II] emission contaminating the
H$\alpha$+[N II]  image near the center. Keel (1983) identified
NGC 4314 as  having an ``[N II]''-type  optical spectrum.
Measured values of [N II]/H$\alpha$ for the NGC  4314 nucleus
range  between 1 and 2.5 (Keel 1983; Wakamatsu \& Nishida 1980; 
Stauffer 1982; Smith  et al.\ 1987) and Keel (1983) identified
NGC 4314 as an ``[N II]''-type galaxy.  The  ratios of H$\alpha$,
H$\beta$, [N II] $\lambda$6584, [O III] 
$\lambda$5007, and [O I] $\lambda$6300 given in Smith et  al.\
(1987) are consistent with the LINER  definition given in Heckman
(1980) \&  Dahari (1985). Therefore the nuclear peaks in 
H$\alpha$+[N II] and (H$\alpha$+[N II])/CO may  not be due to
star formation.

Note the coincidence of an increase in $\mu_{\rm I-J}$ profile, a
decrease in 
$\mu_{\rm B}$, and a  peak in the $\sigma_{\rm H_2}$ profile to
the SE. These all argue that CO and  dust are associated.  How
good is this correlation? Is there CO only where there is dust?
The peak in  the $\sigma_{\rm H_2}$ profile to the SE (Fig.\ 7)
is well-described  by a gaussian centered at $r =  -5\farcs4$,
with FWHM $= 4\farcs14$. From the {\it HST}  
$\mu_{\rm I}$ data presented in P2, we  find the  dust lane can
also be described by a gaussian centered at $r = - 5\farcs5$,
with FWHM 
$= 1\farcs60$.  Along this p.a.\ the OVRO synthesized beam can be
considered to be a  gaussian with FWHM $= 2\farcs20$. Convolving
the OVRO beam with the dust profile  produces a distribution with
FWHM $= 2\farcs70$. Evidently, the CO is not confined to  the
dust lane, but spills over in a symmetric distribution. We next
deconvolve the  OVRO beam from the Fig.\ 7  $\sigma_{\rm  H_2}$
profile and determine that the actual CO  distribution has FWHM
$= 3\farcs51$ and a peak $\sigma_{\rm H_2} = 2550$ ${\rm
M_{\sun}}$ pc$^{-2}$ at the dust lane   center.
	
Along a slice perpendicular to the primary stellar bar (Fig.\ 8),
we again  find strong CO near the nucleus and near the dust lanes
associated with the ring.  (H$\alpha$+[N II])  peaks outside the
CO to the NE, but inside to the SW. The   (H$\alpha$+[N II])/CO
ratio picks up near $r = -4^{\prime \prime}$  at a gap in the CO
ring.  There is an  increase in the SFR at $r= -1^{\prime
\prime}$ , due to the observed CO  deficiency, or to [N II] 
contamination from the LINER phenomenon. To the SW, the redder
dust  lane at 
$r = 6^{\prime \prime}$  is present in CO as well. The CO maximum
at $r \sim  4^{\prime \prime} 5$ to the NE  is not as  pronounced
in the color indices as that to the SW, to be expected if the SW
side  of the galaxy is nearer.  

For completeness we provide a set of profiles (Fig.\ 9) passing
East-West  through knot A, the second most intense H\,{\sc ii}
region seen in the Pogge (1989) data.  The dust to CO correlation
is not as striking, primarily because dust signatures  are weak
in $\mu_{\rm I-J}$ and $\mu_{\rm B-H}$. 

There is one other location where comparisons with optical data
might  prove useful; at the ``dust bowl'' in the primary stellar
bar to the SE at 
$r \sim 23^{\prime \prime}$ .    In Fig.\ 10 we plot profiles
passing through the  dust bowl, perpendicular to the  primary
stellar bar. The profiles are $\mu_{\rm B}$, 
$\mu_{\rm V-I}$, and $\mu_{\rm B-V}$  (P1) along with
$\sigma_{\rm H_2}$  derived from the natural  weight map (Fig.\
3).   Again, dust and CO are well  correlated, with reddening in
$\mu_{\rm V-I}$  matching increases in $\sigma_{\rm H_2}$ all
along the profile.
	
We test the reality of this dust and CO correlation in Fig.\ 11, 
showing the  relationship between the measured CO intensity and
A$_{\rm V}$ across the dust  bowl in  the range $1 \leq \times
\leq 7^{\prime \prime}$ . The A$_{\rm V}$ values were  derived in
P1 from B-V surface  color  indices. The correlation demonstrates
the quality of the CO data at the dust bowl  location. This
suggests that most of the weak CO features detected along the 
primary stellar bar (Fig.\ 3) are probably real.
	  
As a final check of the reality of this feature in CO, we derive
an ${\rm N}_{\rm  p}$ to A$_{\rm V}$  relationship. From Fig.\ 11
we determine
\begin{quote} 
 I$_{\rm CO}$(Jy arcsec$^{-2}$  km s$^{-1})$ = a + b A$_{\rm V}$
(mag) 
\end{quote}  
where $a = 0.080 \pm 0.023$ and $b= 1.955 \pm 0.178$.
Converting the slope to K km s$^{-1}$, using (Table 2) 1 Jy
arcsec$^{-2} = 1.19$ K, we  find  a slope of 2.54 K km s$^{-1}$
mag$^{-1}$. From Bloemen et al.\ (1986) 
\begin{displaymath}
 {\rm N}({\rm H}_2)({\rm cm}^{-2}) = 2.8 \times 10^{20}\; {\rm
I_{\rm CO}}\;  ({\rm K\, km\, s}^{-1}) 
\end{displaymath} 
Hence, 
\begin{displaymath}  {\rm N}({\rm H}_{2}) = 7.1 \times 10^{20}\;
{\rm A_{\rm V}}
\end{displaymath}  
and, introducing a factor of two in converting
from H$_{2}$ to N$_{\rm p}$,
\begin{displaymath}  {\rm N}_{\rm p} = 1.4 \times 10^{21}\; {\rm
A_{\rm V}}
\end{displaymath}  
in general agreement with a value
\begin{displaymath}  {\rm N}_{\rm p} = 1.8 \times 10^{21}\; {\rm
A_{\rm V}}
\end{displaymath}
 derived from Savage \& Mathis (1979), assuming A$_{\rm
V}$/E(B-V)$= 3.1$. The  agreement suggests that, at least in this
part of NGC~4314, a dust screen  description is valid (Witt et
al.\ 1992).

Unfortunately, as discussed in P1, the dust bowl is the only
region in NGC~4314 for which we can obtain reasonable values for
A$_{\rm V}$. This is possible only  by  exploiting the symmetry
and the relatively uniform stellar population of the  primary
stellar bar. In the more complex circumnuclear region, B-V color 
variations are due to spatial variations in the stellar
populations, as well as dust  extinction.

\section{Kinematical Results}

\subsection{The CO Velocity Field and the Rotation Curve}

In Fig.\ 12 we show the naturally weighted velocity field, which
is  consistent with predominately circular motion for $r <
10^{\prime \prime}$ . From these  data, we  have determined the
systemic velocity V$_{\rm sys}$, the dynamical center of the 
galaxy  ($\alpha, \delta$), the position angle of the line of
nodes $\phi$, the inclination 
$i$, and a rotation  curve V(R). We used the iterative technique
of Puche, Carignan, and Wainscoat  (1991), which employs a tilted
ring model. Our data was binned into seven rings,  each 1\farcs5
in width. Velocity values within each ring were weighted by the
cosine  of the azimuthal angle (relative to the line of nodes) in
the plane of the galaxy.  We first held $\phi$ and $i$ constant
and fit for 
$\alpha$,
$\delta$, and V$_{\rm sys}$. Next, we held 
$\alpha$, $\delta$, and  V$_{\rm sys}$ constant and fit for 
$\phi$, $i$, and V(r), fitting the receding and approaching 
halves of the velocity field separately. We included only data
within 45\deg\ of the  major axis. Our results are listed in
Table 2.  
	
The inclination derived from this analysis (21\deg $\pm$
20\deg) is quite uncertain due  to the nearly face-on
orientation of NGC~4314, however it is consistent with the  value
found in P1 (23\deg $\pm$ 8\deg), assuming that  the outer
isophotes obtained from  digitally stacking the POSS O and E
plates are intrinsically circular. However, the  outer isophotes
may be intrinsically oval, since outer rings and pseudorings 
located at the OLR are generally elongated (Buta 1993). An
argument based on  the Tully-Fisher relationship (see Kenney et
al.\ 1993) gives $27\deg \pm 4\deg$. We choose  $i  =
21$\deg\ to  determine the CO-derived rotation curve.

In an effort to extend the rotation curve outward from the OVRO
limit, we  have adopted the rotation curve of Combes et al.\
(1992)  for $r > 10^{\prime \prime}$ .  This curve  is obtained
from a mass model of the galaxy based on a K-band image, which 
presumably accurately traces the stellar mass in the galaxy. We
choose the  Combes model derived rotation curve over that from
Quillen et al.\ (1994), since  the agreement with the CO
velocities in the center of the galaxy is better.

Fig.\ 13a shows our OVRO and the Combes et al. (1992) rotation
curves.   Agreement between the two rotation curves is good for
$r \leq 7^{\prime 
\prime}$ . The CO curve  flattens at $r = 7^{\prime \prime}$ 
with V$_{\rm max} \sim 175$ km s$^{-1}$.    Assuming that the 
ring is co-planar  with the plane of the galaxy, has a radius of
6\farcs7, and a circular velocity of 175  km s$^{-1}$, the period
of rotation for the gas in the ring is 11 My.

\subsection{Deviations from Circular Velocity}

Deviations from circular velocity can be seen directly on the
velocity map  shown in Fig.\ 12. There are regions associated
with the dust lanes just outside the  nuclear ring (e.g., Fig.\
5) where the isovelocity contours bend sharply by
$\sim 90$\deg.  To study these regions in more detail, we
obtain a velocity residual map by  producing a model galaxy
velocity field from the rotation curve, Fig.\ 13a, and 
subtracting it from the observed CO velocity field. This is shown
in Fig.\ 14,  overlaid on the CO intensity map. 

This comparison shows that the largest deviations from circular
motion  occur in two symmetrically located spurs, just outside
the ring and near where  the dust lanes merge with the ring.
These spurs are located along the minor axis,  where, within a
disk, non-circular motions imply radial motions. If the dust
lanes  are truly along the leading edge of the bar, and the
spiral arms are trailing, then  the southwest side of the galaxy
is closest to us and these radial deviations imply  inflow. These
local inflow speeds reach their maximum of 90 km~s$^{-1}$ at $r
\sim  10^{\prime \prime}$  and gradually decrease as the nuclear ring is
approached and crossed. Inflow  along the leading edge of a bar
is predicted by theory and models (e.g., Roberts  et al.\ 1979,
Athanassoula 1992b) but is rarely seen as clearly as it is in
NGC~4314.
	
We can estimate the mass inflow rate into the ring.  The CO flux
in the two  inflow spurs ($\pm 7^{\prime \prime}$ from the nucleus) is 50
Jy km s$^{-1}$, which  corresponds to a gas  mass of M(H$_{2}$ +
He) $= 7 \times 10^7$ M$_{\sun}$.   Adopting a mean inflow speed
of 30 km  s$^{-1}$   implies that the gas now at $7^{\prime \prime}$ will
reach the nucleus in 336 pc/30 km s$^{-1}  =  10^7$ yr.  Thus the
average inflow rate over this time is $7 \times 10^7$
M$_{\sun}$/10$^7$ yr $= 7$ M$_{\sun}$  yr$^{- 1}$.  
	
Strictly speaking, the inflow speeds measured in Figure 14 are 
local  streaming motions, and do not necessarily represent net
inflow speeds.  Gas  along bars has the largest inward motion
where it piles up and is easiest  to detect,  but is predicted to
move outwards on other parts of its orbit, where the gas  surface
density is lower (Athanassoula 1992a).  This is a greater concern
far out  along bars than just outside the ILR ring,  and we
believe that the net inflow  speeds are close to the local
streaming motions in this part of NGC~4314.  
	
From the thermal radio continuum flux, G-B estimate an
extinction-corrected  H$\alpha$ flux for NGC~4314 of $1.5 \times
10^{-12}$ erg s$^{-1}$ cm$^{-2}$,  which    corresponds  to a
star formation rate of 0.16 M$_{\sun}$ yr$^{-1}$, using an
extended Miller- Scalo IMF as  in Kennicutt (1983). Since this
rate is much lower than the inflow rate, the gas  mass  in the
ring may be increasing with time. And, since little CO is seen in
the  dust lanes, this process cannot continue much longer.

In Fig.\ 15b the velocity deviation map is compared to the Pogge
(1989) H$\alpha$  image. Luminous H\,{\sc ii} complexes are
located within 1$^{\prime \prime}$ of the terminus of each  inflow spur.
This suggests that gas flowing inward along the bar collides with
gas  already in the ring, triggering star formation. The H\,{\sc
ii} regions appear upstream  from the inflow zones, suggesting
that it is gas already in the ring that is involved  with star
formation. These regions may be similar to the bar end in M83,
where  two gas streams meet and one of them undergoes star
formation (Kenney \& Lord  1991).
	
The existence of CO near the nucleus implies the need for a
collection  mechanism. In P2 we presented evidence for the
existence of an $8^{\prime \prime}$ long nuclear  bar, oriented parallel to
the primary stellar bar. If this nuclear bar is a real bar-like
dynamical feature, then gas flow patterns of the type thought to
be associated  with bars should be present.  Fig.\ 16 shows that
deviations from circular motion  (the same differential
velocities presented in Fig.\ 14, but with $\Delta V = 5$ km
s$^{-1}$) are  seen along the nuclear bar.  However, the velocity
pattern is not everywhere  consistent with the streaming motions
generally associated with bars. The signal-to-noise ratio in this
central region, where CO emission is weak, may be too low  to
show the streaming motions clearly.

\subsection{Resonances}

From the rotation curve in Fig.\ 13a, we produce standard
resonance curves  (c.f.\ Devereux et al.\ 1992), and plot them in
Fig.\ 13b.  Because the angular  velocity of this bar is unknown,
we assume a pattern speed  of $\Omega_{p} = 3.5$  km s$^{-1}$ 
arcsec$^{-1}$, which places corotation at the end of the primary
bar. A $\mu_{\rm B-I}$ map of  NGC~4314 (Fig.\ 16) shows that it
has an outer  elliptical distribution of newer  stars with $r =
13^{\prime \prime}$ oriented perpendicular to the primary stellar bar
(P1). NGC~4314 also has a nuclear bar with a $4^{\prime \prime}$ radius
(P2) oriented parallel to the primary  stellar bar. These
orthogonal structures suggest nested resonances (Combes  1988).
The resonance curves in Fig.\ 13b would predict OILR at $r = 13$
and IILR  at $r = 4^{\prime \prime}$ for a bar pattern speed $\Omega_{p}
= 3.5$ km s$^{-1}$ arcsec$^{-1}$,  corresponding to 72   km
s$^{-1}$ kpc$^{-1}$. This  places the outer Lindblad resonance
(OLR) at $r \sim  98^{\prime \prime}$,  coincident with the elliptical
outer isophotes discussed in P1. These elliptical  isophotes have
their major axis oriented perpendicular to the primary stellar
bar,  as do some rings and pseudo-rings in other galaxies that
are located at the OLR  (Buta \& Crocker 1993).

\section{Comparisons with other Galaxies}

A concise review placing NGC~4314 in the context of other barred 
galaxies with similar observations  can be found in Kenney
(1995). Here we offer  a few specific comparisons.  In the
central regions of M101 (Kenney et al.\  1992),  NGC~6951 (Kenney
et al.\ 1992),  NGC 1530 (Regan, Vogel, \& Teuben  1995),  and
M100  (Rand 1995; Sakamoto et al.\ 1995),  CO maps show 
spiral-shaped features extending out from  a circumnuclear ring
or partial  ring of H\,{\sc ii}  regions.  In many of these
examples  the CO is strongest at the intersection of the  ring
with the dust lanes along the leading edge of the bar.  This
morphology is  particularly striking in NGC~3351,  where the CO
is concentrated into ``twin  peaks'' located symmetrically  about
the nucleus oriented perpendicular to  the  large-scale bar
(Kenney et al.\ 1992). These peaks have been interpreted as 
regions of orbit crowding, where inflowing gas from the bar meets
gas on  more circular orbits near the ILR (Kenney et al.\ 1992).

Our data for NGC~4314 show spiral-shaped inflow spurs, but they
are  weaker  than in most of these other galaxies.  NGC~4314 also
does not have a  ``twin peak'' morphology.  There are modest CO
peaks located where the  inflow spurs meet  the ring, but the
strongest CO peaks are located elsewhere in  the ring.   The
variety in gas morphology between galaxies  is probably due in 
part to dynamical differences, including  variations in the
degree of  central  concentration and the strength of  the bar. 
The rotation curve and  the  pattern speed of the stellar bar 
determine the number, location, and  separation  of the ILRs. 
Gas behavior also depends on  whether there are small  scale
nuclear bars  (Shlosman et al.\ 1989; Wozniak et al.\ 1995),  and
the relative  position angles of the bars,  if they are
independently rotating.  Some of the  variety may also be due to
evolutionary differences.  Timescales are relatively  short in
circumnuclear regions; starburst timescales are $\sim 10^7-10^8$
yr, and  dynamical timescales  are $\sim10^7$ yr.  The relative
weakness of the CO in the  inflow spurs  of NGC~4314 compared to
these other galaxies may mean that in  NGC~4314
 the gas in these dust lanes has  already been driven into the
center.  	 The observation  that the molecular gas in NGC~4314 
lies inside the star  forming ring  lead Combes et al.\ (1992) to
suggest that the ring is shrinking with  time due to dynamical
friction from the bulge stars.  As noted previously, there is  an
elliptical distribution of newer, but non-ionizing stars outside
of the ring of  H\,{\sc ii} regions (P1).  There is a color
progression along  this ellipse, such that stars  closer to the
ring are bluer (P1).  Since our OVRO data show very little CO 
along this ellipse, we conclude that this color variation is due
to age rather than  extinction.  Therefore, younger  stars are
found closer to the H\,{\sc ii} region ring,  supporting  the
idea of a shrinking ring of star formation.  This observation, 
along with the observation that essentially all of the molecular 
gas in the galaxy  is found in or inside the ring  and the fact
that NGC~4314 is extremely deficient  in atomic gas (G-B), lead 
Combes et al.\ (1992) to hypothesize that NGC~4314  is in a late
stage of evolution, and that accretion to the NGC~4314 ring has 
stopped.  Our observations show that inflow is in fact still
occuring.  However,  since little CO is seen in the dust lanes, 
this process cannot continue much  longer. 

\section{Summary}
	
The major observational results of this paper include
\begin{enumerate}

\item We confirm the ring-like CO morphology first reported by
Combes et  al.\ (1992). 
	
\item We discover significant CO near the center of NGC~4314
without  associated recent star formation.

\item CO associated with dust along the leading edge of the
primary stellar bar  is detected at a 3-$\sigma$ level.  This
association is confirmed for the dust bowl region  by a strong
correlation of CO intensity with extinction estimated from
optical  color maps.

\item Much of the CO in and near the ring is coincident with dust.

\item We identify the nested, orthogonal resonances in NGC~4314.
We  identify the IILR with the end of the nuclear bar, OILR with
the outer extent of  an elliptical distribution of less recent
star formation, corotation near the end of  the large-scale bar, 
and the OLR with the elliptical outer isophotes oriented 
perpendicular to the bar. The strongest star formation occurs
between the IILR  and OILR and closer to the IILR.

\item A color gradient, from blue near the IILR to red near the
OILR, is  likely due to stellar age differences, not reddening.
CO emission is strong near  the IILR and weak near the OILR,
hence the expected reddening gradient is the  opposite of the
observed color gradient.

\item Radial inflow at speeds of 20-90 km s$^{-1}$ is detected in
two spurs of  molecular gas located just outside the star-forming
ring. Significant H$\alpha$ emission  occurs where the inflow
spurs meet the ring.
\end{enumerate}	
\acknowledgements Benedict thanks Jeff Achtermann, Antonio
Garcia-Barreto,  Jim Higdon,  and Hong Bae Ann for useful
discussions. Melody Brayton provided invaluable assistance with
LaTex. We thank Rick Pogge for providing an  H$\alpha$ map
of NGC~4314. Benedict acknowledges support from  NASA grant 
NAG5-1603. B.~J.\ Smith acknowledges support from NASA grant
NAG2-67.

\clearpage

\section*{Figure Captions}

\begin{figure}[htbp]
\caption{An unsharp masked image of NGC 4314 from P1. This
high-pass map  emphasizes regions within the galaxy where
luminosity gradients are changing  rapidly. It brings into
prominence dust lanes (white) and (for $r \sim 6$\arcsec)
confirmed  regions of recent star formation (black). The sharp
gradient of the nucleus also  shows up as black. Note the
$26\arcsec$ diameter oval-shaped distribution at p.a.\ $\sim
58$\deg,  and the dusty region (the ``dust bowl'') along the
bar, southeast of the nuclear ring  at RA = 12$^{\rm h}\, 22^{\rm
m}\, 32\fs7$,  Dec = +29\deg\ 53\arcmin\ 27\arcsec\ (2000).}
\end{figure}
\begin{figure}[htbp]
\caption{The uniformly weighted channel maps. The velocity
interval is 13 km  s$^{-1}$. Channel 9 corresponds to $V =
1075.5$ km s$^{-1}$, while channel 24 is $V = 880.5$  km
s$^{-1}$. The contour levels are 3, 4, ... , 10 times our uniform
map noise level of 28  mJy beam$^{-1}$. The dynamical center is
marked with a cross in each channel.}
\end{figure}
\begin{figure}[htbp]
\caption{The uniform weight CO contours superposed on the
gray scale CO  map. The beam size is $2\farcs3
\times 2\farcs2$. The contour levels are 0.1, 0.2, ... , 1.0
times the  peak intensity of 1.3 Jy beam$^{-1}$ km s$^{-1}$. The
3$\sigma$ level is 0.153 Jy beam$^{-1}$ km s$^{-1}$,  slightly
higher than the lowest plotted contour.}
\end{figure}
\begin{figure}[htbp]
\caption{The natural weight CO contours superposed on the
gray scale CO  map. The beam size is $3\farcs1
\times 2\farcs7$. The contour levels are 0.1, 0.2, ... , 1.0
times the  peak intensity of 2.74 Jy beam$^{-1}$ km s$^{-1}$. The
3$\sigma$ level is 0.165 Jy beam$^{-1}$ km s$^{-1}$,  slightly
lower than the lowest plotted contour. The extended distribution 
lies along  the primary stellar bar at p.a.\ = 148\deg.}
\end{figure}
\begin{figure}[htbp]
\caption{	a) A uniform weight CO intensity contour map
superposed on the  central region of Fig.\ 1.  The second lowest
contour level is 5$\sigma$ above the noise  level. The gray scale
is encoded such that local excess (newer stars) is black and 
local deficit (dust) is white. Note that the elliptical locus of
newer stars oriented  along p.a.\ = 58\deg\ is devoid of CO.
b) A natural weight CO contour map (Fig.\ 4)  superposed on Fig.\
1.  Note coincidence of CO with the leading edges of the primary
stellar bar and with  the dust bowl.}
 \end{figure}
\begin{figure}[htbp]
\caption{The uniform weight CO contour map from Fig.\ 3
superposed on an  H$\alpha$ + [N II] gray scale map (Pogge 1989).
The gray scale is encoded to represent the  strongest  H$\alpha$
emission as black. Map registration was determined by
identifying  the Wakamatsu \& Nishida (1980) H$\alpha$-bright
knots A and B.}
 \end{figure}
\begin{figure}[htbp]
\caption{Detailed profiles along p.a.\ = 148\deg, the
major axis of the primary bar.  Surface magnitude and color
indices ($\mu_{\rm B}$, $\mu_{\rm B-H}$,  and $\mu_{\rm I-J}$
are from P1.  The surface  density of CO is from the uniform
weighted map (Fig.\ 3).  The transformation of  CO to molecular
gas surface densities is described in the text. We model the gas 
surface density peak to the SE  with a gaussian. The overlay
shows the range and  quality of the fit. $\mu_{{\rm H}\alpha}$ is in arbitrary
units. In the bottom panel, the ratio  H$\alpha$/$\sigma_{\rm
H_2}$  provides a qualitative and relative indication of the
local SFR.  The missing data in  $\mu_{\rm I-J}$ and $\mu_{\rm
B-H}$ near the nucleus are due to point spread function
differences between  the visible and short wavelength infrared
(SWIR) bandpasses (see P1).}
\end{figure}
\begin{figure}[htbp]
\caption{Same as Fig.\ 7, except along p.a.\ = 58\deg, 
the minor axis of the primary  bar.}
 \end{figure}
\begin{figure}[htbp]
\caption{Same as Fig.\ 7, except along p.a.\ = 0\deg,
through the H\,{\sc II} region, knot A.}
 \end{figure}
\begin{figure}[htbp]
\caption{Profiles through the dust bowl perpendicular to
the primary bar. The  $\mu_{\rm B}$, $\mu_{\rm B-V}$  and
$\mu_{\rm V-I}$ data are from P1. There are no J, H or K data for 
this region.}
\end{figure}
\begin{figure}[htbp]
\caption{Visual absorption (A$_{\rm V}$) and CO surface
density ($M_{\sun}$  pc$^{-2}$) are  correlated for the gas and
dust at the dust bowl.}
\end{figure}
\begin{figure}[htbp]
\caption{The naturally weighted velocity field superposed
on a uniformly  weighted gray scale map. The contour interval is
$\Delta V = 20$ km s$^{-1}$. We have labeled  several contours.
The broken circle contour  to the NW is $V = 1040$ km s$^{-1}$.
The  similar contour to the SE is $V = 940$ km s$^{-1}$.}
\end{figure}
\begin{figure}[htbp]
\caption{ a) The rotation curve from OVRO and Combes et
al.\ (1992) data. The OVRO data extend out only to 9\arcsec.  The
Combes data extend the rotation curve to $r  > 120$\arcsec. The
errors in {\it V} are $1 - \sigma$.  The error bars in radius represent
the beam size.  b)  The resonance curves. A pattern speed of  3.5
km s$^{-1}$ arcsec$^{-1}$ places the  IILR at the end of the
nuclear bar (P2) and the OILR coincident with the outer  ellipse
of newer star formation discussed in P1 and shown in Fig.\ 14.}
\end{figure}
\begin{figure}[htbp]
\caption{Velocity residuals produced by subtracting the
Fig.\ 13 rotation curve  from the observed natural weight
velocity field in Fig.\ 12. Residuals are  superposed upon the
Fig.\ 3 uniform weight CO intensity gray scale map. The  velocity
interval is 10 km s$^{-1}$.}
\end{figure}
\begin{figure}[htbp]
\caption{a) The natural weight CO velocity residual map
superposed on the  optical unsharp masked data.  b) The natural
weight CO velocity residual map superposed on  H$\alpha$+[N II]
from Pogge (1989). The gray scale is encoded to represent the
strongest  H$\alpha$+[N II] emission as black. Note that the
deepest incursions of the decelerating  gas terminate in close
proximity to large H\,{\sc ii} regions.}
\end{figure}
\begin{figure}[htbp]
\caption{Surface I-band intensity (greyscale) from HST (P2)
and CO velocity  deviations from purely circular motion
(contours). $\Delta V = 5$ km s$^{-1}$ . The nuclear bar  is
aligned along p.a\. = 148\deg.}
\end{figure}
\begin{figure}[htbp]
\caption{B-I surface color ($\mu_{\rm B-I}$) map of NGC
4314. Encoded such that darker  to lighter represents bluer to
redder. Note the discrete steps in the color progression  along
the feature identified with the OILR.}
\end{figure}

\clearpage

\section*{Tables}

{\sc TABLE} 1. Global Properties of NGC 4314

{\sc TABLE} 2. OVRO Interferometer Results for NGC 4314

\end{document}